\documentclass[twocolumn]{aastex63}

\def\lapp{\ifmmode\stackrel{<}{_{\sim}}\else$\stackrel{<}{_{\sim}}$\fi}
\def\gapp{\ifmmode\stackrel{>}{_{\sim}}\else$\stackrel{>}{_{\sim}}$\fi}
\usepackage{multirow}
\usepackage{color}
\usepackage{amsmath}
\usepackage{soul}
\usepackage{hyperref}

\shorttitle{}
\shortauthors{}
\begin{document}

\title{Spectral variability of the blazar 3C~279 in the optical to X-ray band during 2009--2018}

\correspondingauthor{Hongjun An}
\email{hjan@cbnu.ac.kr}

\author{Sungmin Yoo and Hongjun An}
\affiliation{Department of Astronomy and Space Science, Chungbuk National University, Cheongju, 28644, Republic of Korea}

\begin{abstract}
	We report on spectral variability of the blazar 3C~279
in the optical to X-ray band between MJD~55100 and 58400 during which
long-term radio variability was observed.
We construct light curves and band spectra in each of the
optical ($2\times10^{14}$--$1.5\times10^{15}$\,Hz) and X-ray (0.3--10\,keV) bands,
measure the spectral parameters (flux $F$ and spectral index $\alpha$),
and investigate correlation between $F$ and $\alpha$ within and across the bands.
We find that the correlation of the optical properties dramatically
change after $\sim$MJD~55500 and the light curves show more frequent
activity after $\sim$MJD~57700.
We therefore divide the time interval into three ``states'' based on the correlation properties and
source activity in the light curves, and analyze each of the three states separately.
We find various correlations between the spectral parameters in the states
and an intriguing 65-day delay of the optical emission
with respect to the X-ray one in state~2 (MJD~55500--57700).
We attempt to explain these findings using a one-zone synchro-Compton emission scenario.
\end{abstract}


\section{Introduction}
\label{sec:intro}
        Blazars, the most energetic radiation sources in the Universe,
are active galactic nuclei (AGNs) with one of the jets
pointing toward Earth \citep[][]{up95}.
Their large energy output is believed to be produced in the central
region by rapid spin of a supermassive black hole \citep[][]{bz77}
and flows outwards in the form of bipolar jets.
The relativistic particles in the jets produce radiation which is further
boosted due to Doppler beaming, and so blazars are bright across
the entire electromagnetic wavebands.

        As bright blazars can be seen even at very
high redshifts \citep[$z>5$;][]{rsgp04}, they are very useful to
study environments in the early Universe and its Cosmic evolution
\citep[e.g.,][]{hessebl13}. Furthermore,
blazars are energetically favorable sources of
ultra high-energy Cosmic rays (UHECRs) and neutrinos, and
can give us important clues to the acceleration mechanisms of the
$>10^{15}$\,eV particles \citep[e.g.,][]{rfgb+18}.
These studies require detailed knowledge on the spectral energy
distribution (SED) of blazars' emission for making beaming correction
\citep[e.g.,][]{ar18}, characterizing absorption by the extragalactic
background light \citep[EBL; e.g.,][]{aaa+16}, and constraining the jet contents
\citep[e.g.,][]{brsp13}.

        Blazars' emission SEDs are phenomenologically well characterized by double-hump
structure: a low-energy hump in the optical to X-ray band and
a high-energy one in the X-ray to gamma-ray band.
The low-energy hump is believed to be produced by synchrotron radiation
of electrons, and the high-energy one by inverse-Compton (IC) upscattering
of internal (synchrotron-self-Compton; SSC) or external (external Compton; EC)
soft-photon fields \citep[e.g.,][]{dermer95}.
It was also suggested that additional hadronic contributions could
be important in some blazars \citep[e.g.,][]{brsp13,bbpc16}.
Note that low-frequency radio photons
are self-absorbed (synchrotron-self-absorption; SSA) in the compact
high-energy emitting jets, and so the observed radio photons
are believed to be emitted further downstream of the jet in these models.

\begin{table*}[t!]
\caption{Observational data used in this work}
\label{ta:ta1}
\centering
\scriptsize{
\begin{tabular}{lcl}
\hline
Instrument        & Band & Refs. \\ \hline
OVRO & 15\,GHz & https://www.astro.caltech.edu/ovroblazars/ \\
{\it WISE} & Bands 1--4  & https://www.nasa.gov/mission\_pages/WISE/main/index.html \\
Steward & VR & http://james.as.arizona.edu/$\sim$psmith/Fermi/ \\
SMARTS & BVRJ  &  http://www.astro.yale.edu/smarts/glast/home.php \\
{\it Swift}/UVOT & 170--600\,$nm$  & https://swift.gsfc.nasa.gov/ \\
{\it Swift}/XRT & 0.3--10\,keV & https://swift.gsfc.nasa.gov/ \\ \hline
\end{tabular}}
\end{table*}

        This one-zone picture cannot explain all the diverse phenomena observed
in blazars' emission but captures main features of the blazar SEDs.
More complicated models \citep[e.g.,][]{mmjj15}
were also developed and applied to some blazars with limited success.
Nevertheless, particle acceleration mechanisms, structure of the jet flow,
and the composition of the jets are not yet very well known.
Because the emission mechanisms for the two SED humps differ,
the frequency and time dependence of their
variability induced by jet activities can give
us crucial information on the jet structure and particle acceleration mechanisms.
These have been studied by SED modeling and multi-wavelength
variability analyses \citep[e.g.,][]{pss15,lrfk+18}.

        3C~279 is a very bright and highly variable blazar
\citep[$z=0.536$;][]{msdc+96},
and is categorized as a flat-spectrum radio quasar (FSRQ). It exhibits
complex multi-wavelength variabilities: long-term (years) radio variability,
short-term (days) optical and X-ray flares, and minute-scale gamma-ray
flares \citep[e.g.,][]{hnms+15}. Some of these flares show correlation
in multiple wavebands \citep[e.g.,][]{pfcl+18,bdam19,ljmv+20,prince20}. As such,
3C~279 can give us insights into blazar jet physics with its rich temporal and spectral properties.
In this paper,
we present our spectral variability studies performed using $\sim$9-yr observations
in the optical to X-ray band.

\section{Data reduction and Analysis\label{sec:sec2}}

In order to construct band spectra in each of the
optical ($2\times 10^{14}$--$1.5\times 10^{15}$\,Hz) and
X-ray (0.3--10\,keV) bands, high-cadence nearly contemporaneous multi-frequency
data are needed. We therefore analyze data taken with
the {\it Neil-Gehrels-Swift} satellite, and supplement these with
data taken from public catalogs.
The data used in this work are listed in Table~\ref{ta:ta1}.

\subsection{Swift data analysis\label{sec:sec2_1}}
        Since 3C~279 was monitored frequently with the Neil-Gehrels-{\it Swift}
observatory, it provides relatively high-cadence
data in the X-ray (0.3--10\,keV; XRT) and six optical bands (170--600\,$nm$; UVOT).
We download the observational data in the HEASARC data archive
and use the UVOT and XRT data for our studies.

        For the UVOT data analysis, we use an $R=5''$ circular and
an $R=20-30''$ annular regions centered at 3C~279 for the source and background,
respectively. We then measure the source flux
using the {\tt uvotsource} tool integrated in HEASOFT~v6.22.

\begin{figure*}
\centering
\includegraphics[width=160mm]{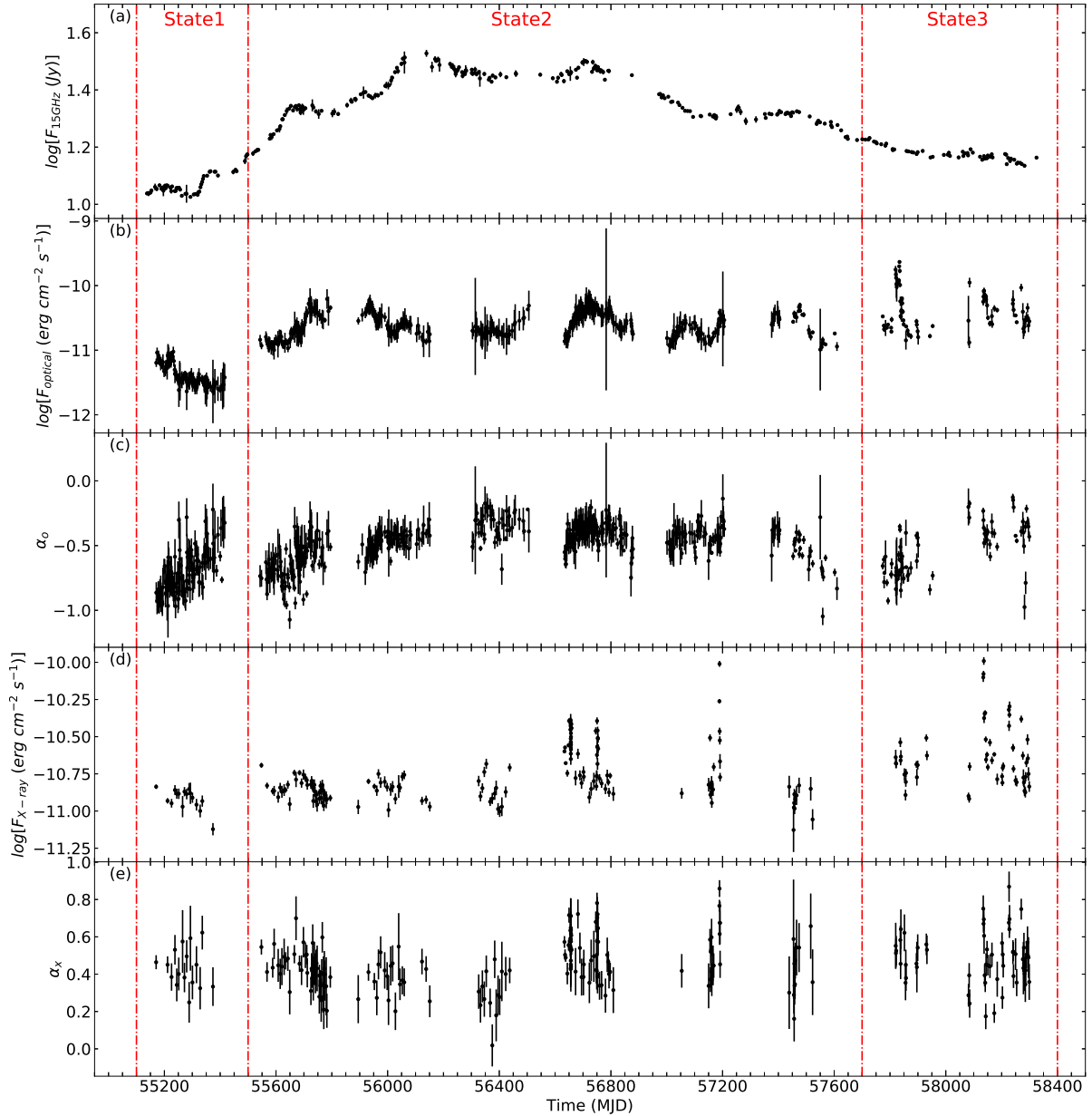}
\figcaption{Multi-frequency light curves of 3C~279: ({\it a}): OVRO 15\,GHz flux,
({\it b}): optical flux, ({\it c}): optical spectral index,
({\it d}): X-ray flux, and ({\it e}): X-ray spectral index.
Vertical lines denote the time intervals for the states 1, 2, and 3 (see text), and
the radio light curve is shown for reference.
\label{fig:fig1}}
\end{figure*}

        For the XRT data, we first reprocess the data using
{\tt xrtpipeline} to produce cleaned event files.
We then perform a spectral analysis using $R=R_{\rm in}-70''$ (with $R_{\rm in}$
varying depending on the degree of pile up)
and $R=120-210''$ annular regions for the source and background, respectively.
Note that photon pile-up occurred in some of the observations because of X-ray flares
of 3C~279 \citep[see also][]{ljmv+20}. In this case, we further inspect the event distribution
and excise central regions affected by pile-up.\footnote{https://www.swift.ac.uk/analysis/xrt/pileup.php}
The corresponding ancillary files are produced with {\tt xrtmkarf}, and we
use pre-computed redistribution matrix files (RMFs).
We then fit the spectra with power-law models ({\tt tbabs*pow}) holding $N_{\rm H}$ fixed at
$2.2\times 10^{20}\rm \ cm^{-2}$ in {\tt XSPEC}~12.9.1p to measure
the source flux and produce the X-ray light curve.
For the absorption model, we use {\tt vern} cross section \citep[][]{vfky96}
and {\tt angr} abundance \citep[][]{angr89}. The fits are well acceptable with
the typical $\chi^2$/dof=0.98.
The resulting light curve is shown in Figure~\ref{fig:fig1}.
Note that the {\it Swift} data used in this work were also analyzed and presented
previously \citep[e.g.,][]{ljmv+20, prince20}, but we carry out more detailed
`spectral' studies in this paper.

\subsection{Public catalog data\label{sec:sec2_2}}
        {\it Swift} UVOT provides optical data with reasonable quality but the cadence
is insufficient for our studies; other data are necessary to cover the gaps.
So we supplement the UVOT data with the public
SMARTS- and Steward-catalog ones \citep[Table~\ref{ta:ta1};][]{smrt+09,bubb+12}.
We correct the optical data for Galactic
extinction using $A_\lambda$ values found in
NED\footnote{https://ned.ipac.caltech.edu/extinction\_calculator} \citep[][]{sf11},
convert the magnitude into flux units \citep[][]{bcp98} and generate light curves.
Note that some of the data were also presented
in previous works \citep[e.g.,][]{pss15,pfcl+18,ljmv+20,prince20}.
We also use limited {\it WISE} observations ($\sim$$10^{13}$--$10^{14}$\,Hz) here;
these are not used for band spectral fits but for qualitative characterization of the SEDs.
We also show the 15\,GHz radio light curve obtained in the OVRO catalog \citep[][]{rmpk+11}
for reference (Fig.~\ref{fig:fig1} a).

\subsection{Time-series SED fitting\label{sec:sec2_3}}
        Although the narrow-band fluxes (i.e., each observation)
provided important insights into blazar jets previously
\citep[e.g.,][]{ljmv+20,prince20},
variability in the spectral shape cannot be studied in details with this approach.
Since spectral shapes can be measured only with multi-frequency data,
we combine the data together to construct time series of
the spectra in the optical and X-ray bands.

        For each of these bands, we combine the data within one day and construct SEDs
in each of the wavebands.
Using slightly different time bins (e.g., 2--3\,days)
does not significantly alter the results presented below.
Note that when we compare quantities measured by {\it Swift}/XRT,
we do not bin the data in time.

        We fit the optical SEDs with power-law models $K(\nu/\nu_0)^\alpha$, where $\nu_0$
is the pivot frequency taken to be the geometric mean of the fit band.
We then measure the ``logarithmic'' flux ($F_o$; ``flux'' hereafter)
and the spectral index ($\alpha_o$). In order to ensure that the observational data
cover a wide frequency range ($\nu_{\rm min}$--$\nu_{\rm max}$) in the fits,
we require that $\mathrm{log_{10}}(\nu_{\rm max}/\nu_{\rm min})$
is greater than 0.4 for the optical data.
This requirement does not have large impact on the fits
since the observations cover the fit band well.
We verify that the models reasonably
represent the SEDs by visual inspection.

	The optical-band fits are formally unacceptable with reduced $\chi^2_r\gg1$, meaning
that the measurement uncertainties are underestimated and/or the simple power law
is inadequate to fit the high-quality optical data; these will make the uncertainties
on the model parameters incorrectly small. Although it is unclear what
the poor fits should be ascribed to, 
we increase the measurement uncertainties by a factor so as to make
the fit reduced $\chi_r^2=1$, which also increases the uncertainties in $F_o$ and $\alpha_o$.
The results are displayed in Figure~\ref{fig:fig1} (b and c).
We note that the Pearson correlation coefficient and its Fisher transformation
\citep[][]{fisher1915} we use below take into account the scatter
in the data (e.g., $F_o$ and $\alpha_o$ measurements), and so the measurement
uncertainties are indirectly accounted for via the scatter.
We verify the results obtained from the Fisher transformation
using simulations when necessary (e.g., \S~\ref{sec:sec3_1}).

        The $0.3-10$\,keV X-ray data are separately
fit in {\tt XSPEC} (see \S\ref{sec:sec2_1}),
and we measure the logarithmic flux ($F_x$) and
derive the SED slope ($\alpha_x=2-\Gamma_X$).
We present the results in Figure~\ref{fig:fig1} (d and e).
The {\it WISE} data cover three $\Delta T\approx 1$\,day epochs at MJDs~55205, 55379,
and 55567, and reveal that the optical continuum SED might curve downwards
below $\sim$$10^{14}$\,Hz at some epochs.
The measured SED slopes are $-0.49\pm0.03$/$-0.78\pm0.13$,
$-0.59\pm0.03$/$-0.59\pm0.12$, and $-0.72\pm 0.04$/$-0.77\pm0.25$ for the {\it WISE}/optical data
at MJDs~55205, 55379, and 55567, respectively.

\section{Optical-to-X-ray variability of 3C~279 \label{sec:sec3}}
        The light curves in Figure~\ref{fig:fig1} show various phenomena at the
observed frequencies. Long-term (years) and short-term (months)
variabilities are clearly seen in the light curves. Some flares are observed
only in one passband (e.g., the X-ray flare at MJD~56750 and the optical flare at
MJD~57830), while some others are observed in multiple wavebands (e.g., $>$MJD~58000).
It is hard to explain all these observational diversities with an emission scenario, and
thus we focus on some of the features and a one-zone scenario here
\citep[see also][]{pss15,ljmv+20,prince20}.

\begin{table}[t!]
\caption{Summary of correlations between the spectral properties}
\label{ta:ta2}
\centering
\scriptsize{
\begin{tabular}{lcccccc}
\hline
Band1   &   Band2  & Property     & $r_p$   & Sig. ($\sigma_F$) & $N_{\rm pair}$ & \\ \hline
\multicolumn{6}{l}{Full data (MJD~55100--58400):}  \\
Optical &  Optical & $F_o$/$\alpha_o$ & 0.44 & 12.4 & 693 & \\
X-ray   &  X-ray   & $F_x$/$\alpha_x$ & 0.65 & 12.0 & 239 & \\
Optical &  X-ray   & $F_o$/$F_x$      & 0.48 & 7.0 & 175 & \\
Optical &  X-ray   & $\alpha_o$/$\alpha_x$    & $-$0.04 & 0.5 & 175 & \\ \hline
\multicolumn{6}{l}{State~1 (MJD~55100--55500):}  \\
Optical &  Optical & $F_o$/$\alpha_o$ & $-$0.61 & 7.7 & 121 & \\
X-ray   &  X-ray   & $F_x$/$\alpha_x$ & 0.14 & 0.5 & 16 & \\
Optical &  X-ray   & $F_o$/$F_x$      & 0.25  & 0.9  & 16 & \\
Optical &  X-ray   & $\alpha_o$/$\alpha_x$    & 0.04  & 0.1 & 16 & \\ \hline
\multicolumn{6}{l}{State~2 (MJD~55500--57700):}  \\
Optical &  Optical & $F_o$/$\alpha_o$ & 0.41  & 9.5 & 490 & \\
X-ray   &  X-ray   & $F_x$/$\alpha_x$ & 0.72  & 11.5 & 166 & \\
Optical &  X-ray   & $F_o$/$F_x$      & 0.18  & 1.9 & 115 & \\
Optical &  X-ray   & $\alpha_o$/$\alpha_x$    & $-$0.08  & 0.9  & 115 & \\ \hline
\multicolumn{6}{l}{State~3 (MJD~57700--58400):}  \\
Optical &  Optical & $F_o$/$\alpha_o$ & 0.07  & 0.6 & 82 & \\
X-ray   &  X-ray   & $F_x$/$\alpha_x$ & 0.59  & 5.0 & 57 & \\
Optical &  X-ray   & $F_o$/$F_x$      & 0.54  & 3.9  & 44 & \\
Optical &  X-ray   & $\alpha_o$/$\alpha_x$  & $-$0.07 & 0.5 & 44 & \\ \hline
\end{tabular}}
\end{table}

        In one-zone blazar emission models,
the low-energy (optical) and the high-energy (X-ray to gamma-ray) radiations
are related as they are assumed to share the emitting particles (i.e., electrons)
in the same region \citep[][]{fermipol10}.
Hence, correlation between optical and X-ray emission should
exist, and we search the spectral data for such correlation.
Here, we consider two spectral properties, flux ($F_i$) and spectral index ($\alpha_i$)
in the optical ($i=o$) and X-ray bands ($i=x$), which makes up four correlations:
one correlation between the spectral properties in each band (2 total) and two cross-band
correlations for each property (2 total).
Note that this approach is slightly different from the previous ones
\citep[e.g.,][]{pfcl+18,bdam19,ljmv+20, prince20}
in the sense that we use spectral indices as well and measure fluxes over
broader bands which represent the continuum better.

\begin{figure*}[t]
\centering
\includegraphics[width=175mm]{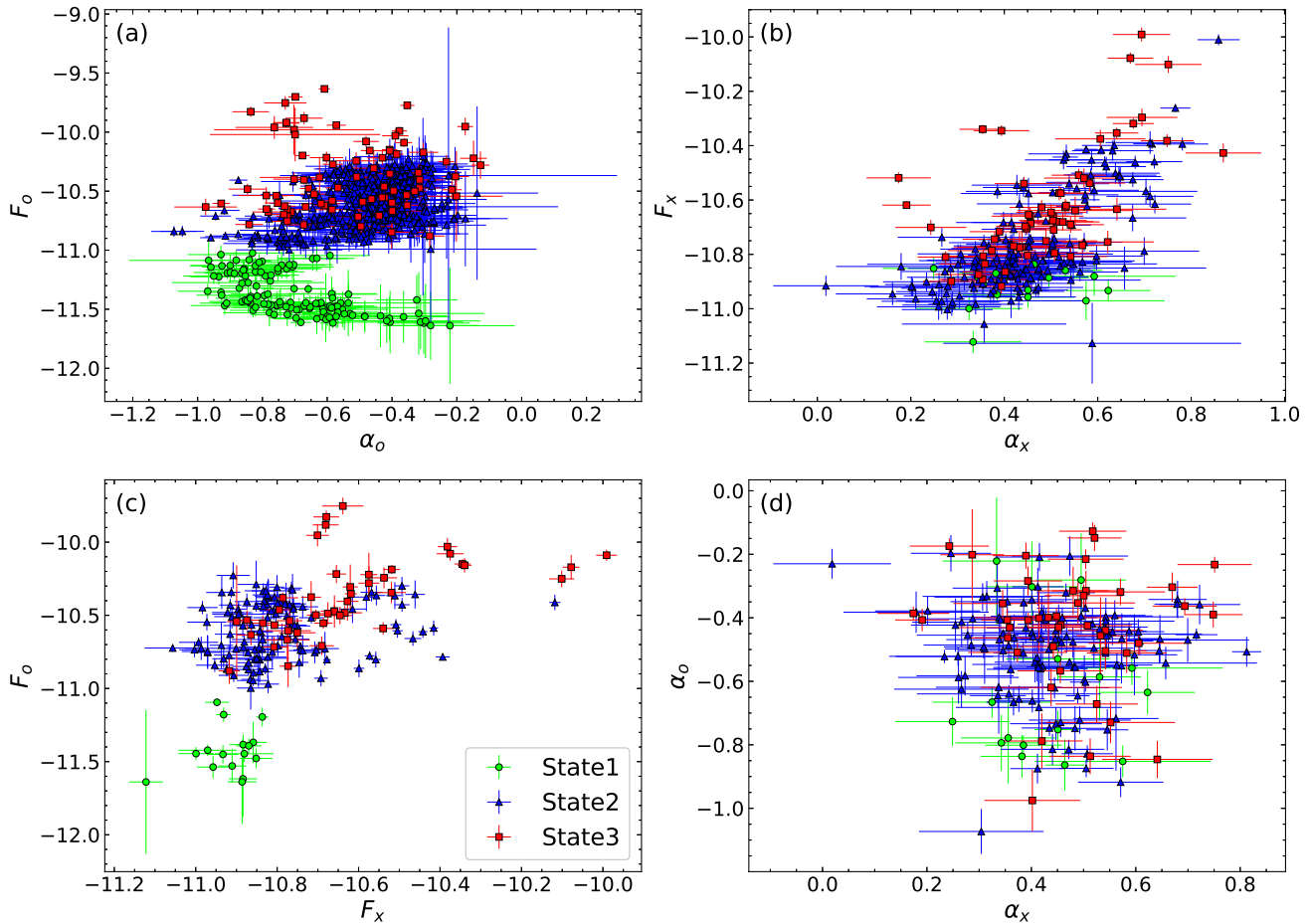}
\figcaption{Scatter plots of the fit fluxes and spectral indices to show correlation within
and across the wavebands. Correlations between $F_i$ and $\alpha_i$
within a waveband are shown in the top row: optical (a) and X-ray (b) bands.
Cross-band correlations between the fit fluxes (c: $F_o$--$F_x$)
and between the spectral indices (d: $\alpha_o$--$\alpha_x$) are displayed
in the bottom row. Data points for each state are denoted in
color: green circle for state~1, blue triangle for state~2 and red square for state~3.
\label{fig:fig2}}
\end{figure*}

\subsection{Correlations within and across the wavebands \label{sec:sec3_1}}
        With the band-fit fluxes $F_i$'s and spectral indices $\alpha_i$'s
that we measured (\S\ref{sec:sec2_3}), we calculate the Pearson correlation
coefficients ($r_p$) for the four pairs. The significance ($\sigma_F$) for the correlation
is computed with the Fisher transformation.  The results are summarized
in Table~\ref{ta:ta2} and scatter plots are shown in Figure~\ref{fig:fig2}.
We verify that the significance estimated by the Fisher transformation well represents
the null hypothesis probability using simulations; i.e., the chance probabilities
for uncorrelated random samples (drawn from the normal distribution) to show the $r_p$
values in Table~\ref{ta:ta2} correspond to $\sigma_F$'s estimated by the Fisher transformation.

	In the single-band correlation study, we find very significant (e.g., $\ge$5$\sigma$)
correlations in both bands (9-yr data).
Although the optical flux $F_o$ and spectral index $\alpha_o$ show very significant correlation
over the 9-yr period, it appears that there are two different ``states''
with dramatically different trends
(i.e., negative and positive correlation) with a boundary
at $F_o\approx -11$ (Fig.~\ref{fig:fig2} a).

        In the cross-band correlation study, we find
significant correlations in $F_o$--$F_x$ \citep[Fig.~\ref{fig:fig2} c; see also][]{ljmv+20}.
Like the $F_o$--$\alpha_o$ case (see above), the $F_o$--$F_x$ relation
appears to form two groups depending on the $F_o$ values (Fig.~\ref{fig:fig2} c).
In addition, light curves after MJD~57700 show more frequent activities at high energies,
differing from the earlier ones. We therefore group the data into three states:
(1) MJD~55100--55500 with low optical flux, (2) MJD~55500--57700 with high optical flux
and mild activity, and (3) MJD~57700--58400 with high optical flux and strong activity.
Note that the time intervals for these states are similar to those used by \citet{ljmv+20}
based on the $R$-band and gamma-ray flux relations, but are slightly different from theirs
in that we do not use earlier data ($<$MJD~55100) and our state 3 includes their
intervals~3 and 4.

        The results for correlation studies in the states~1--3 are presented in
Table~\ref{ta:ta2}.  Note that the correlation in the $F_o$--$F_x$
relation (Fig.~\ref{fig:fig2} c) in the full data set almost disappears in states 1 and 2,
and is weaker in state 3, meaning that the correlation in the full data set is primarily
between the states (inter-state) rather than within them (intra-state) and that
the correlation in state 3 is an intra-state one.
While the inter-state correlations can tell us about
the state transition of the blazar, we focus on the intra-state ones here.

\begin{figure*}[t]
\centering
\includegraphics[width=160mm]{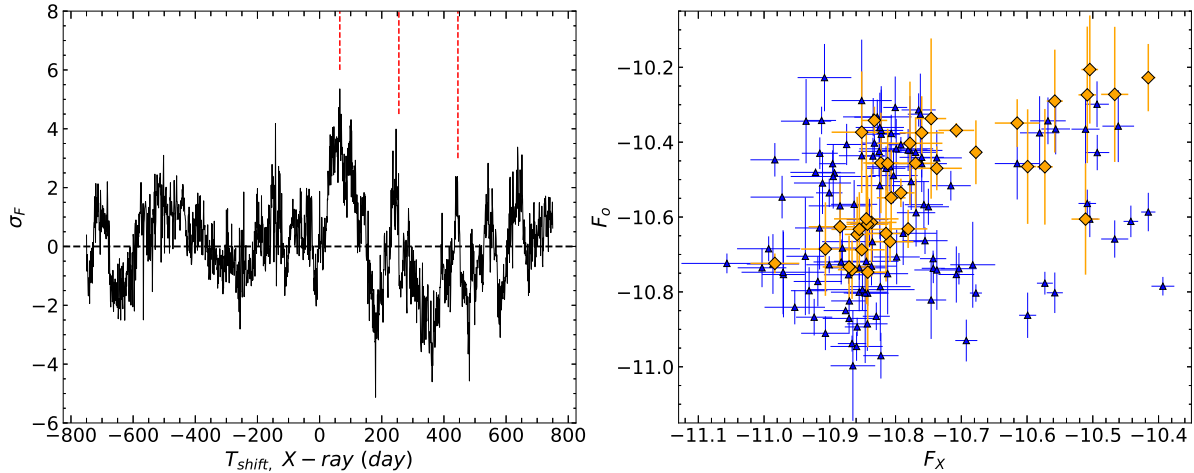}
\figcaption{{\it Left}: significance plot for time-shift cross-band correlations between
$F_o$ and $F_x$ in state~2. The x axis $T_{\rm shift}$ denotes the time shifts
(1-day steps) and the shifted data
with the positive values for X-ray lead.
Negative y values mean anti-correlation. Red vertical lines mark
65\,day, 255\,day, and 445\,day.
{\it Right}: the corresponding scatter plot with $F_x$ shifted by 65\,days (orange diamond).
Unshifted data are overlaid in blue triangle for reference.
\label{fig:fig3}}
\end{figure*}

\subsection{Time-shifted Correlations \label{sec:sec3_2}}
        Because the cross-band correlation may be more significant with a time delay
if emission in one band lags (leads)
the other, we perform the same correlation study by shifting the data in time. This
is essentially the same as the discrete correlation function \citep[DCF;][for example]{lrfk+18}
method except that our data are binned.
We compare cross-band properties in pairs of the two wavebands by shifting one of the
data in time and measure the correlation significance ($\sigma_F$)
as a function of the time shift ($T_{\rm shift}$; 1-day step).
The results are consistent with those in Table~\ref{ta:ta2}; the time-shifted plots
corresponding to the significant (cross-band) ones in the table show a prominent
peak at $T_{\rm shift}=0$.

	However, we find that $F_o$ and $F_x$ in state 2,
whose correlation was insignificant without a time shift (Table~\ref{ta:ta2}),
show significant correlation when one of them is shifted in time
($r_p=0.73$ and $\sigma_F\approx$5.4 at $\sim$65\,days; Fig.~\ref{fig:fig3} left).
As noted above (\S~\ref{sec:sec3_1}), $\sigma_F$ well represents the
null hypothesis probability, and so the false alarm probability
for the correlation with the 65-day delay (pre-trial $\sigma_F=5.4$) is $p=6\times 10^{-5}$
after considering 1,500 trials (i.e., $\sim$4$\sigma$ post-trial).
A similar delay of 64\,day found in
MJD~56400--56850 by an independent study \citep[][]{pfcl+18} enhances the significance for the
shifted correlation.
In our new analysis of the data, we find that this delayed
correlation in state~2 (MJD~55500--57700) is stronger over the whole period of
the state than in a part of it \citep[e.g., MJD~56400--56850;][]{pfcl+18}, implying that the delay
persisted for a longer period (e.g., the whole state).
We also note that Figure~\ref{fig:fig3} seems to show a possible periodic trend
with a period of 190\,days (red vertical lines in the left panel).
This is intriguing, but significance for the later peaks is low.
The periodic trend in the figure might appear just by chance.

        For the time-shifted $F_o$--$F_x$ correlation in state~2, we show the scatter plots
of the shifted (orange diamond) and unshifted data (blue triangle) in Figure~\ref{fig:fig3} right.
In this state, there were several
X-ray flares (Fig.~\ref{fig:fig1}) which can be seen as high-flux outliers
in Figure~\ref{fig:fig3} right. Since the source would have different emission
properties during the flare periods from the low-flux quiescent ones \citep[e.g.,][]{hnms+15},
we check to see if the delayed correlation exists in the quiescent and flare periods separately.
Because of the paucity of data points in flare, we investigate the quiescent periods only.
We remove the high-flux outliers (i.e., taking $F_x \le -10.75$), and compute $r_p$ and its
significance which are $\approx0.65$ and $\sigma_F\approx 4$, respectively.

\section{Interpretation of the Spectral Correlation using a toy SED model \label{sec:sec4}}
        In this section, we present our explanation on the observed variabilities
using a simple one-zone scenario.

\begin{figure*}[t]
\centering
\begin{tabular}{cc}
\includegraphics[width=80mm]{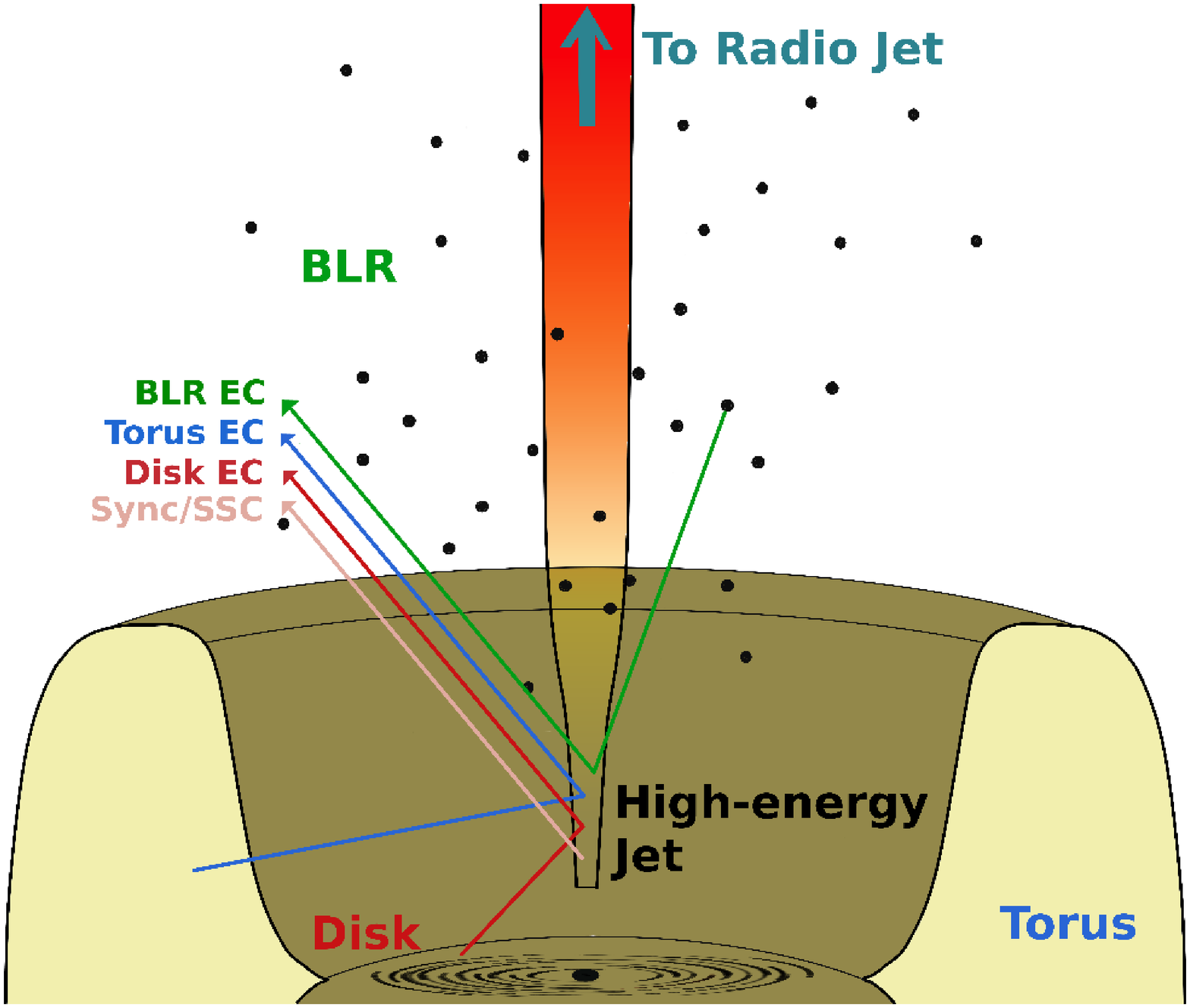} &
\includegraphics[width=80mm]{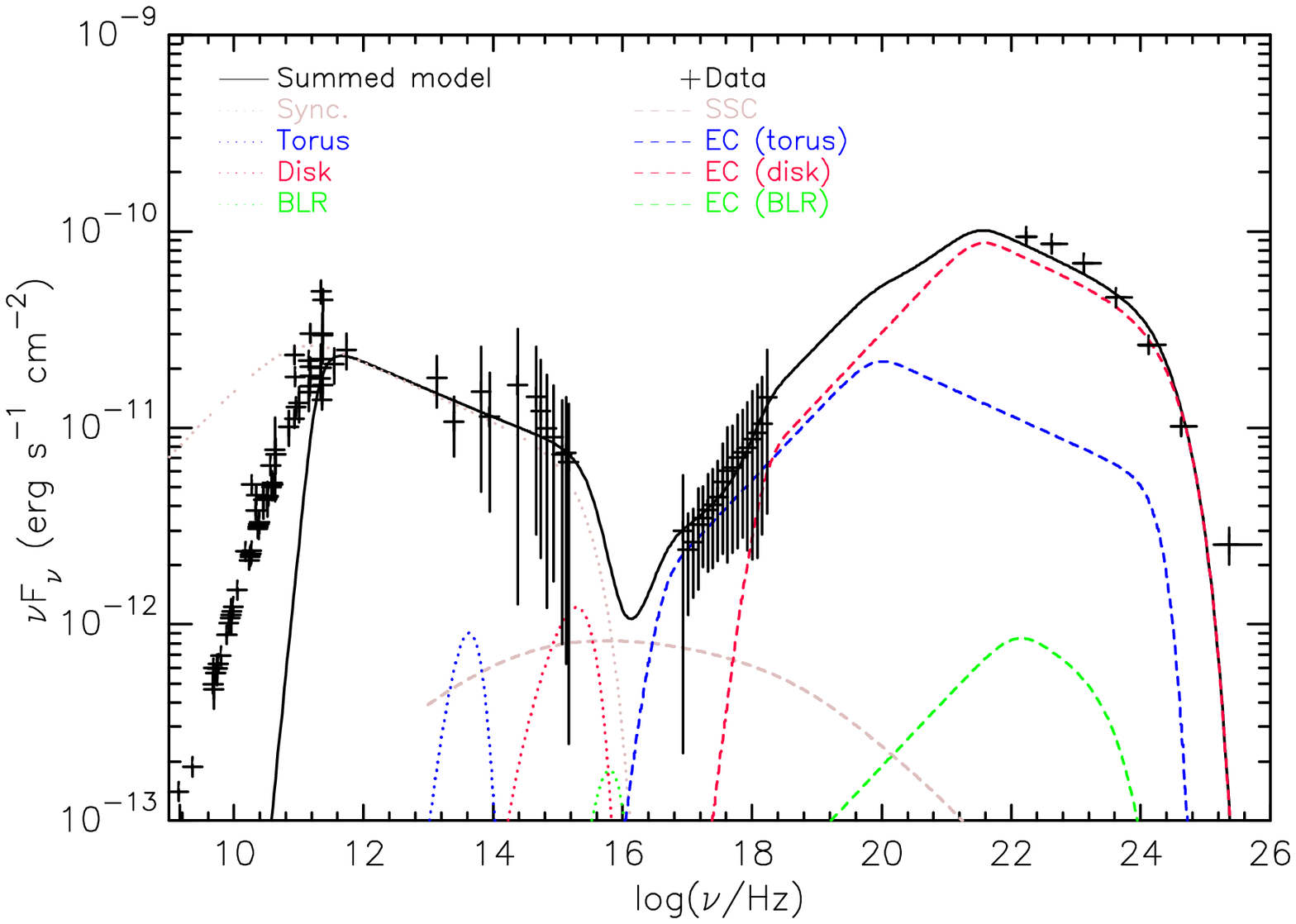} \\
\end{tabular}
\figcaption{A schematic view of blazars' emission components
(left; not to scale), and
a toy one-zone SED model and time-averaged SED data (right).
The optical and X-ray data points are measured in this work (collected over $\sim$9~years),
and the error bars on the data points are the standard deviation of the measurements.
The radio and gamma-ray points are taken from the {\tt vizier} catalog and the 4FGL
catalog (8-year average), respectively. The summed model is displayed with a solid black line and
each model component is denoted in color.
Note that the synchrotron component (brown dotted) is shown without SSA to be compared to
a model with SSA (black solid).
\label{fig:fig4}}
\end{figure*}

\subsection{A Toy SED model \label{sec:sec4_0}}

        In order to explain the spectral correlations we found above,
we construct a toy one-zone SED model using a leptonic
synchro-Compton scenario \citep[Fig.~\ref{fig:fig4}; see also][for example]{sg12,dclb+14,pss15}.
In this work, we do not attempt to strictly match the highly-variable observational SEDs of 3C~279 with
the toy model, but it is constructed so as to capture the main features of the SED and the relevant
ingredients in blazar emission for our investigation of the spectral correlations;
the figure is intended to be used only to guide eyes.
The model and time-averaged SED data are shown in Figure~\ref{fig:fig4}, where
the error bars on the optical and X-ray data points are standard deviation
of the 9-yr flux measurements at the observed frequencies.
Note that the gamma-ray SED is obtained from the 4FGL catalog \citep[][]{fermi4fgl}
and is a mission-averaged one, and the radio data are taken from the {\tt vizier} photometry
webpage.\footnote{http://vizier.unistra.fr/vizier/sed/}

        In this model, the optical SED is explained with
the synchrotron radiation (brown dotted) of a broken power-law electron distribution
($dN_e/d\gamma_e\propto \gamma_e^{-p_1}$ with $p_1=3.3$ if $\gamma_e\ge50$
and $p_1=2.3$ otherwise) and weak disk emission at the high-frequency end (red dotted).
The spectral indices for the electron distribution are chosen so as to match the optical
SED displayed in Figure~\ref{fig:fig4} right and are similar to those expected
in shock acceleration theories \citep[e.g.,][]{je91} and radiative cooling.
The disk emission is computed following the standard Shakura-Sunyaev model
for $M_{\rm BH}=5\times 10^8 M_\odot$ \citep[][]{ss73}.
Emissions of a torus and a broad line region (BLR) are included as blackbody radiation
\citep[e.g.,][]{jmb14};
our investigation below does not strongly depend on the exact emission properties of these
components (e.g., the emission frequency and spectral shape).
We also show the SSC component (brown dashed) for reference but it may be even lower;
given the observed shape of the average X-ray SED, this component cannot be significant.
In the X-ray band, the emission is assumed to be produced by EC of torus (blue dashed) and
disk photons (red dashed).
Although the BLR EC is much weaker than the disk EC in our model, the converse
is also possible if the jet locates closer to BLR/torus \citep[e.g.,][]{dermer95};
BLR emitting at slightly lower frequencies may replace the disk EC in the model.

        Variability in this scenario can occur for various reasons: changes
in internal conditions of the jet
(e.g., particle spectrum $N_e$, magnetic-field strength $B$, the Doppler factor $\delta$),
location of the jet (i.e., EC efficiency), and
change in the external seeds for EC (e.g., variable external emission $u_{\rm ext}$).
In the model, the frequencies and fluxes of the low-energy (synchrotron in the optical band) and the
high-energy (EC in the X-ray to gamma-ray band) SED humps are related to the jet
properties as in the following \citep[e.g.,][]{dermer95,ar17}:
\begin{eqnarray*}
\nu_{\rm SY} & \propto & \delta B \gamma_e^2,\mathrm \ \ \ \ \ \  \ 
F_{\rm SY} \propto B^{(1+p_1)/2} \delta^{(5+p_1)/2} \\
\nu_{\rm EC} & \propto & \delta^2 \gamma_e^2 \nu_{\rm ext},\mathrm \ \ \ 
F_{\rm EC} \propto \delta^{3 + p_1} u_{\rm ext},
\end{eqnarray*}
where $\nu_{\rm SY, EC}$ and
$F_{\rm SY,EC}$ are the observed (synchrotron and EC) emission frequency and flux,
and $\nu_{\rm ext}$ and $u_{\rm ext}$
are the emission frequency and flux of the external seeds for EC.
Given the relatively straight power-law shape of the optical spectrum
of 3C~279 (Fig.~\ref{fig:fig4}), the optical spectral index would not change much by
$\delta$ and/or $B$ (shifts of the SED)
unless they vary a lot (e.g., orders of magnitude), in which case the flux would change
even more. So variation of the optical spectral index ($\alpha_o$)
would be likely due to changes of
the particle spectrum $N_e$ with small contribution from $\delta$ and/or $B$.

        Note that this model accounts only for the high-energy jet in which radio emission
is highly suppressed by SSA (black solid vs. brown dotted lines).
The radio emission is assumed to be produced in a separate parsec-scale `radio' jet.

\subsection{State~1: MJD~55100--55500 \label{sec:sec4_1}}
        In state~1 during which the optical flux is low, $F_o$--$\alpha_o$
shows negative correlation. No cross-band correlation is found.
Although the changes of $F_o$ and $\alpha_o$ may occur for various
reasons as we noted above, the `negative' $F_o$--$\alpha_o$ correlation may suggest
that the change is stronger at low frequencies (i.e., soft), and can occur
if $N_e$ varies in the jet region.

        A change of $N_e$ would necessarily result in a corresponding change in the X-ray SED
in one-zone scenarios; the optical and X-ray
light curves (Fig.~\ref{fig:fig1}) show a hint of a correlated flux change (both drop
with time), but the correlation is not significantly detected, perhaps because of
the low statistics (16 pairs in this state). We verify this using simulations
performed with correlated random samples and with the measured data; only $\le$1$\sigma$
detection of $r_p\approx0.25$ correlation (e.g., $F_o$/$F_x$ in Table~\ref{ta:ta2}) is possible with 16 pairs.
Note that gamma-ray flux also drops in this state \citep[e.g.,][]{ljmv+20}.
These imply that the soft-spectrum particles ($N_e$) were being removed from the high-energy
emission region in this state. We may speculate that these particles move to the radio jets,
thereby producing radio emission; the radio brightening in
state~1 (Fig.~\ref{fig:fig1}) may support our speculation
although the radio activity may be irrelevant to the optical one and
was produced by a shock propagating in the radio
jet \citep[][]{hntv+08} and/or independent changes in conditions in the radio jet: $\delta$, $B$,
and/or injection of particles.

\subsection{State~2: MJD~55500--57700 \label{sec:sec4_2}}
        In this state, properties within the bands are all positively correlated.
In particular, the $F_x$-$\alpha_x$ correlation can give us strong constraints on the
emission mechanism of the high-energy ($\ge$X-rays) radiation.
This state overlaps very well with interval 2 of \citet[][]{ljmv+20} in which
the $R$-band ($F_R$) and gamma-ray ($F_\gamma$) flux relation
of $F_\gamma\propto F_R^{7.7}$ is found.
We note that the significant $F_o$--$\alpha_o$ correlation is primarily due to
grouping of low-flux (e.g., $F_o \le -10.8$)
and high-flux points (blue points in Fig.~\ref{fig:fig2} a);
ignoring the low-flux ones reduces the significance rapidly.
This indicates that state~2 may be further split, but
the low-flux points do not localize in time.
Therefore, we do not further split this state and regard that
the $F_o$--$\alpha_o$ correlation is less significant
(e.g., $\sim$3.8$\sigma$ for $F_o\ge -10.8$).

        Aside from the radio variability,
observational properties in this state are that
(1) $F_\gamma\propto F_R^{7.7}$,
(2) $F_x$ and $\alpha_x$ show ``positively'' correlated variability, implying that the
disk EC varies more than the torus EC does,
(3) $F_x$ leads $F_o$ by $\sim$65 days (Fig.~\ref{fig:fig3} left), and
(4) variability in the optical spectral index implies $N_e$ variation.
Because $F_{\rm EC}/F_{\rm SY}\propto \delta^{\frac{1+p_1}{2}}u_{\rm ext}/B^{\frac{1+p_1}{2}}$,
changes of $\delta$, $B$ (by a factor of $\le 3$), and/or $u_{\rm ext}$ can explain (1). However,
(2) is hard to be produced by the internal properties $\delta$ and $B$
(e.g., $\alpha_x$ variability), and therefore we can conclude
that $u_{\rm ext}$ is the primary source of the X-ray and hence gamma-ray variability.
Furthermore, (3) cannot be explained by changes in the internal properties
of the jet either because these will change the X-ray emission instantaneously (i.e., no delay).

        Although the enhanced X-ray and gamma-ray emission should be driven by an increase in the external
disk seed photons ($u_{\rm ext}$), the `delayed' optical emission (3) cannot be produced by
the external sources themselves (e.g., disk) whose emission is much weaker than the
synchrotron continuum (e.g., Fig.~\ref{fig:fig4}) and
will `lead' the reprocessed (upscattered) X-rays. Therefore, we speculate
that the disk (external) activity responsible for (1) and (2)
might enhance the optical continuum emission $\sim$65\,days
later by synchrotron radiation in the `jet'; perhaps enhanced injection from the disk
to the jet is responsible for this.

	The `delayed' optical variability (synchrotron)
is natural in this scenario as $N_e$ would vary by injection,
but then the X-ray flux
should also increase simultaneously by EC of the same $N_e$.
Then $F_x$--$F_o$ correlation with no delay in addition to the 65-day shifted one
is expected but we do not see correlation without a delay (e.g., Fig~\ref{fig:fig3} left).
Perhaps, $F_x$ variability induced by the injection into the jet ($N_e$)
is swamped by the 65-day earlier disk EC activity ($u_{\rm ext}$).
Alternatively, the `delayed' optical continuum variability (synchrotron in the jet)
may be driven mainly by changes of $B$ which
affect $F_o$ but not $F_x$. Indeed, the change of the optical flux (a factor of $\sim$4)
is larger than that of the X-ray one (a factor of $\sim$2 ignoring large X-ray flux
points $F_x\ge-10.8$ induced by flares; Fig.~\ref{fig:fig3} right),
suggesting that $B$ may be the dominant factor (over $N_e$) for the delayed optical variability.

\subsection{State~3: MJD~57700--58400 \label{sec:sec4_3}}
        This state shows rapid and large variability in the high-energy band,
suggesting that 3C~279 was active. Since our spectral data do not cover
this time interval well, highly significant (e.g., $\ge$5$\sigma$) correlation
is found only between $F_x$ and $\alpha_x$ (Table~\ref{ta:ta2} and
red square points in Fig.~\ref{fig:fig2} b), implying that the disk EC is still
variable. If the disk EC is the main driver of the high-energy variability,
a strong $F_\gamma$--$F_R$ trend as in state~2 is expected. However,
\citet[][]{ljmv+20} reported a weaker $F_\gamma\propto F_R^{1.9}$ trend in this
state. This implies that the optical continuum emission also varied with
the X-ray one; fairly significant 4-$\sigma$ $F_o$--$F_x$ correlation (Table~\ref{ta:ta2})
also suggests this \citep[see also][]{prince20}.

        Provided that the optical flux is dominated by the synchrotron continuum radiation,
it is likely that changes of $N_e$ and $\delta$ are the main driver
of the variability in this state. Assuming that $B$ is constant and $p_1\approx2.3$,
we find $F_{\rm EC}\propto F_{\rm SY}^{1.5}$, similar to
$F_\gamma\propto F_R^{1.9}$ trend. Hence the variability in this state
can be explained by changes of $\delta$ and $N_e$
(spectral shape change in the optical band) with relatively weak disk EC
variability ($F_x$--$\alpha_x$ correlation).

\section{Summary and Conclusion\label{sec:conclusion}}
        We investigated multi-frequency variability in the emission of
the blazar 3C~279 using {\it Swift} UVOT/XRT, and various
catalog data. We produced multi-band spectra and analyzed them in
order to infer physical properties and emission mechanisms of the jet.
We constructed time-resolved
SEDs in the two wavebands, carried out spectral correlation studies,
and found several significant correlations between the spectral properties
within and across the wavebands. Note that variability in 3C~279 emission
is much more complicated and may differ in each flare activity \citep[e.g.,][]{pss15,hnms+15},
and that the correlations we found represent the overall properties of the 3C~279 jet.

        In these spectral studies, we found that
the $F_o$--$\alpha_o$ correlation exhibits a strong inversion,
and therefore we split the data into three states based on the $F_o$--$\alpha_o$
correlation and activity in the light curves. We then investigated correlation
properties in each state and interpret the results using a one-zone synchro-Compton
scenario. Below are the summary:
\begin{itemize}
\item State~1: the $F_o$--$\alpha_o$ anti-correlation and
a mild flux drop in the optical to gamma-ray band with time
suggest that soft-spectrum particles ($N_e$) are lost from the high-energy jet.

\item State~2: the $F_\gamma\propto F_R^{7.7}$ relation \citep[][]{ljmv+20}
and $F_x$--$\alpha_x$ correlation
imply that the disk-EC emission is variable due to some activity in the disk.
Then the $\sim$65-day lag of $F_o$ with respect to $F_x$ and variability in
the optical spectral index suggest that the activity in the disk might inject
particles $N_e$ and $B$ into the high-energy jet on a time scale of $\sim$65\,days.

\item State~3: $F_x$--$\alpha_x$ correlation again implies variability in
the disk EC, and the $F_o$--$F_x$ correlation and the $F_\gamma\propto F_R^{1.9}$
relation suggest that it is $\delta$ that mainly drives the variability in
this state with some contribution of $N_e$ and disk EC.
\end{itemize}

	If it is the disk EC emission that drives the
variability at X-rays in state~2 as we argued above,
a 65-day delay of the optical emission
with respect to the gamma-ray one is also expected in state~2 and
was seen in the earlier part of the state \citep[][]{pfcl+18}.
This implies that the emission mechanisms for X-rays and gamma rays
are the same as in our SED model (Fig.~\ref{fig:fig4}).

        It is interesting to note that the near ``quiescent'' state (state~1),
a short period ($\sim$400\,days) after large radio activity \citep[e.g.,][]{cjmo+08},
is followed by active states (states 2 and 3) in which injection of $N_e$ and changes of
$B$/$\delta$ occur. The time interval is coincident with new long-term radio activity,
and it is worth investigating whether and how the high-energy ($\ge$optical)
states are related to the radio activity. Investigations of high-energy data
taken during previous and future long-term radio activity may be very intriguing.

        Because 3C~279 is very bright and frequently observed, high-quality multi-band data
exist. Using only small part of the observational data and a one-zone SED scenario,
we were able to suggest that the emission mechanism for the high-energy SED hump
is the tours and disk EC,
and explore causes of the spectral variability.
Although the one-zone model can explain the results
obtained in this work, the source exhibits enormously diverse spectral variability
that cannot be explained with the model.
More data (e.g., time-varying broadband SEDs) and improved SED models can certainly
provide very useful information. In this regard, we acknowledge that our interpretation
is speculative rather than definitive. Further comprehensive data analyses
\citep[e.g., including polarization and gamma-ray emission;][]{fermipol10}
and theoretical studies (e.g., magnetohydrodynamic simulations and multi-zone SED models)
are warranted to advance our knowledge on 3C~279, and blazars in general.

\bigskip
\acknowledgments

We thank the anonymous referee for the careful reading of the paper and
insightful comments.
This research has made use of data from the OVRO 40-m monitoring
program \citep[][]{rmpk+11} which is supported
in part by NASA grants NNX08AW31G, NNX11A043G, and NNX14AQ89G and NSF
grants AST-0808050 and AST-1109911.
This publication makes use of data products from the Wide-field Infrared Survey Explorer,
which is a joint project of the University of California, Los Angeles,
and the Jet Propulsion Laboratory/California Institute of Technology,
funded by the National Aeronautics and Space Administration.
Data from the Steward Observatory spectropolarimetric
monitoring project were used. This program is supported by Fermi Guest Investigator grants
NNX08AW56G, NNX09AU10G, NNX12AO93G, and NNX15AU81G.
This paper has made use of up-to-date SMARTS optical/near-infrared light curves
that are available at www.astro.yale.edu/smarts/glast/home.php.
We acknowledge the use of public data from the {\it Swift} data archive.
This research was supported by Basic Science Research Program through
the National Research Foundation of Korea (NRF)
funded by the Ministry of Science, ICT \& Future Planning (NRF-2017R1C1B2004566).

\vspace{5mm}
\facilities{OVRO, {\it WISE}, Steward, SMARTS, {\it Swift}}
\software{HEAsoft (v6.22; HEASARC 2014), XSPEC \citep[][]{a96}}

\bibliographystyle{apj}
\bibliography{3C279_apj.bib}
\end{document}